\documentclass[aps,a4paper,superscriptaddress,twocolumn]{revtex4}
\usepackage{tensor}
\usepackage{graphicx}
\usepackage{amsmath}
\usepackage{amssymb}
\usepackage{enumerate}
\usepackage{subfigure}
\usepackage{tabularx}
\usepackage[colorlinks=true, pdfstartview=FitV, linkcolor=blue, citecolor=red, urlcolor=black, breaklinks=true]{hyperref}
\newcommand{\be}{\begin{equation}}
\newcommand{\ee}{\end{equation}}
\newcommand{\ben}{\begin{eqnarray}}
\newcommand{\een}{\end{eqnarray}}
\newcommand{\bes}{\begin{subequations}}
\newcommand{\ees}{\end{subequations}}
\def\bal#1\eal{\begin{align}#1\end{align}}
\newcommand{\vphi}{\varphi}

\newcommand{\LL}{{\mathcal L}}

\begin{document}
\preprint{ITEP-TH-19/18}
\title{Planar ringlike vortices}
\author{D. Bazeia}\affiliation{Departamento de F\'\i sica, Universidade Federal da Para\'\i ba, 58051-970 Jo\~ao Pessoa, PB, Brazil}
\author{M.A. Marques}\affiliation{Departamento de F\'\i sica, Universidade Federal da Para\'\i ba, 58051-970 Jo\~ao Pessoa, PB, Brazil}
\author{D. Melnikov}\affiliation{International Institute of Physics, Universidade Federal do Rio Grande do Norte, 59078-970 Natal, RN, Brazil}\affiliation{Institute for Theoretical and Experimental Physics, 117218, B. Cheremushkinskaya 25, Moscow, Russia}
\begin{abstract}
\end{abstract}

\begin{abstract}
We investigate the presence of vortex structures in generalized Maxwell-Higgs and Chern-Simons-Higgs models in the three-dimensional spacetime. Despite the important difference between the Maxwell and Chern-Simons dynamics, we have been able to introduce first order differential equations that solve the equations of motion for static and rotationally symmetric field configurations. In both cases, solutions of the first order equations engender minimum energy, and we have found vortex configurations whose internal structure unveils interesting and unusual ringlike profile.  
\end{abstract}
\maketitle

\section{Introduction}
Topological structures appear in physics in a variety of contexts. They are important, for instance, in high energy physics, where they usually arise as static solutions of the field equations and may play a role in the structure formation in the primordial Universe and in the cosmic evolution \cite{vilenkin,manton,weinberg}. They also may find applications in condensed matter, where they may describe specific behavior of superconductors and magnetic materials \cite{fradkin,hubert}.

Among the several topological objects that appear in nonlinear science, the most known ones are, perhaps, kinks, vortices and monopoles, which come out in $(1,1)$, $(2,1)$ and $(3,1)$ spacetime dimensions, respectively. Kinks are the simplest of them and only require the action of a single real scalar field to be investigated \cite{vachaspati}. To study vortices, one must take a complex scalar field coupled to an Abelian gauge field through a $U(1)$ symmetry \cite{NO,vega}. The magnetic monopoles are yet more intricate, and require the use of an $SU(2)$ symmetry that couples a triplet of scalar fields to a non Abelian gauge field \cite{thooft,polyakov}.

In this work, we concentrate on vortices. Such structures started being investigated in 1858 by Helmholtz in the context of fluid mechanics \cite{helmholtz,fluidmec}. Almost a century later, Abrikosov noticed that these objects may also appear in type II superconductors \cite{abrikosov} in the Ginzburg-Landau theory of superconductivity \cite{glvortex}. However, it was only in the 1973 Nielsen-Olesen paper \cite{NO} that a relativistic model supporting vortex solutions was presented. In this case, the vortex is electrically neutral and presents quantized magnetic flux.

Over the years, these planar structures have been investigated in generalized models, which may give rise to non standard features \cite{g1,g2,g3,g4,g5,g6,g7,g8,compvortex,vtwin,godvortex,anavortex}. For instance, in Refs.~\cite{g1,g2}, some properties of vortices in Chern-Simons models \cite{csjackiw} were simulated through generalizations of the magnetic permeability in Maxwell-Higgs models. Another non canonical result appeared in Ref.~\cite{compvortex}, where a route to compactify the vortex was introduced; interestingly, a model for the compact vortex seemed to model the magnetic field of an infinitely long solenoid. One may also seek for the presence of vortices in generalized models where the dynamics of the gauge fields is controlled by the Chern-Simons term \cite{csbazeia}. In this scenario, the vortex exhibits a nonvanishing electric field and it tends to compactify into a ringlike structure; see Ref.~\cite{cscomp}.

Vortices may also be investigated in models with enlarged symmetries to accommodate other fields and bring new degrees of freedom. This idea was used in the past to study superconducting strings in Ref.~\cite{witten}, with a $U(1)\times U(1)$ symmetry involving two gauge fields and two complex scalar fields that interact through an extension of a potential of the Higgs type. This type of enhanced symmetry can also be useful to include the so called hidden sector which is of current interest since it can be one of the possible origins of dark matter \cite{p1,p2,p3,p4}. The hidden (or dark) sector may be coupled to the visible (or bright) sector through the scalar fields, as in the Higgs portal \cite{hp1,hp2} and/or through the coupling of the gauge field strengths \cite{dm1,dm2,dm3,dm4,dm5} or even under the presence of generalized magnetic permeabilities \cite{hidden}. In particular, in the last case, we have found that the hidden scalar field may control the generalized magnetic permeability of the visible sector and produce vortex configurations in the visible sector endowed with nontrivial internal structure.

Following the direction that introduces symmetry enhancement, some authors have considered interesting possibilities; see, for instance, Refs.~\cite{sh1,sh2,sh3}, where the $U(1)$ symmetry is changed to become $U(1)\times SO(3)$, with the addition of a triplet scalar field, to add extra degrees of freedom to the vortex. Along the same lines, very recently we have considered, in Ref.~\cite{vortexint}, a $U(1)\times Z_2$ symmetry, which was used to add a neutral scalar field with the $Z_2$ symmetry, to drive the generalized magnetic permeability included in the model. In this case, the scalar field played the role of a source field to generate an internal structure to the vortex. Moreover, we have also investigated magnetic monopoles for a model similar to the 't Hooft-Polyakov one \cite{thooft,polyakov} with the $SU(2)\times Z_2$ symmetry; see Ref.~\cite{monopoleint}. Interestingly, the scalar field also led to an internal structure to the monopole in three spatial dimensions.

There are other interesting investigations on vortices in distinct scenarios that appeared in Refs.~\cite{A,B,C} and in references therein. For instance, in \cite{A} the authors found pipeline vortex structures in a gauge model with global $SU(2)$ and local $U(1)$ symmetries, described under the action of two charged scalar fields. The case of a two-component $U(1)\times U(1)$ model was also studied in Ref.~\cite{C}, and in Ref.~\cite{B} the effect of a dark scalar field with Higgs portal coupling and a $U(1)$ gauge field with a kinetic mixing between two gauge fields is investigated on a semilocal string with a global $SU(2)$ and local $U(1)$ symmetries in the visible sector. Evidently, there are several other possibilities to modify the model and investigate specific features of their topological structures.

The presence of vortices with nontrivial structure in the aforementioned enlarged model \cite{vortexint} has motivated us to seek for simpler models that allow for the presence of the same feature, that is, to construct models that support vortices endowed with internal structure. So, in this work we investigate models similar to the ones described in Refs.~\cite{g1,g2,compvortex,csbazeia,cscomp}, with the standard $U(1)$ symmetry. We consider models of two distinct types, one of the Maxwell-Higgs type, and the other controlled by the Chern-Simons dynamics. Although the Maxwell and Chern-Simons dynamics are very different from each other, in both cases the models support first order differential equations that satisfy the equations of motion, which are obtained by using the Bogomol’nyi-Prasad-Sommerfield (BPS) procedure \cite{ps,bogo}.

The work is further motivated by the current interest in vortices in condensed matter, and here we emphasize some issues, as the ones described in Refs.~\cite{CM1,CM2,CM3,CM4,CM5}, and in references therein. For instance, in \cite{CM1,CM2,CM3}, the authors deal with vortices in dipolar Bose-Einstein condensates, in which the interaction between atoms are of the dipole-dipole type, leading to the presence of vortices that behave differently, acquiring nontrivial structures. Moreover, in \cite{CM4,CM5} the intertwining of two superconducting orders may lead to a charge density modulation that is similar to a pair-density-wave in superconducting vortex halos. Another motivation is related to the inclusion of the 
functions of the scalar field multiplying the Maxwell term, or the scalar kinetic term, which often appear in supersymmetric and supergravity extensions, in particular in the context of holography; see, e.g., the study of a holographic superconductor that admits an analytic treatment near the phase transition \cite{HO1}, an insulator model with nonsingular zero temperature infrared geometry \cite{HO2}, and the transport of electric charge in a strongly coupled quark-gluon plasma \cite{HO3}. 

The gain with the presence of first order equations that solve the equations of motion has motivated us to further investigate the subject, so we organize the current study as follows: in Sec.~\ref{mh} we deal with the Maxwell-Higgs model and present the general properties of the model, finding the equations of motion that describe the vortex and the energy-momentum tensor. In order to get first order equations, we use the BPS procedure and also calculate the energy of the vortex. We then develop new examples and investigate the possibility of adding internal structure to the vortices. In Sec.~\ref{cs}, we extend the investigation to the Chern-Simons-Higgs model and introduce a new model that supports internal structure. Interestingly, the two distinct types of models are capable of engendering vortex configurations endowed with the ringlike profile. In the Chern-Simons case, the vortex is charged electrically, and this makes the problem harder to investigate. Anyway, we study an interesting model, in which the corresponding electric field is also enriched with the ringlike feature, behaving in a way similar to the magnetic field. Finally, in Sec.~\ref{conclusions} we close the work with some comments, conclusions and perspectives for future works.

\section{Maxwell-Higgs Models}\label{mh}

Let us start working in $(2,1)$ flat spacetime dimensions with the Lagrangian density
\be\label{lmodel}
\LL = - \frac{P(|\vphi|)}{4}F_{\mu\nu}F^{\mu\nu} +|D_\mu\vphi|^2 - V(|\vphi|).
\ee
Here $\vphi$ is complex scalar field and $|\vphi|^2= \overline{\vphi}\vphi$, with the overline denoting the complex conjugation. $A_\mu$ is the Abelian gauge field, $F_{\mu\nu}=\partial_{\mu}A_\nu-\partial_{\nu}A_\mu$ is the electromagnetic tensor and $D_\mu=\partial_{\mu}+ie A_{\mu}$ stands for the covariant derivative. The potential is $V(|\vphi|)$ and the metric tensor $\eta_{\mu\nu}$ is diagonal, with $(1,-1,-1)$ being its diagonal elements, and $\hbar=c=1$. 

We refer to function $P(|\varphi|)$ as to the generalized magnetic permeability, which is supposed to be a nonnegative function of the scalar
field modulus. This type of interactions naturally appears in the study of generic supersymmetric extensions of gauge theories. We note, however, that the $N=2$ superfield formalism implies that $P$ is real part of a holomorphic function of $\varphi$, while in the present paper we consider $P$ a real function of $|\varphi|$. Taking $P(|\varphi|)=1$ one recovers the standard case, investigated in Refs. [7, 8]. As we have commented in the previous Section, this kind of generalization was considered before in several other contexts, in particular in \cite{g1,g2} to simulate features of vortices of the Chern-Simons system in the above Maxwell-Higgs system, in \cite{HO2} to study an Einstein-Maxwell-dilaton holographic model for an insulator, and also in \cite{HO3}, to study thermal photon and dilepton production in a strongly coupled plasma of quarks and gluons in a holographic model of the Einstein-Maxwell-dilaton type. In the present context the generalized model will give rise to vortex-like solutions that possess internal structure. We expect this to be of interest in the study of vortices in dipolar Bose-Einstein condensates, which are also endowed with internal structure \cite{CM1,CM2,CM3}.

The equations of motion associated to the Lagrangian density \eqref{lmodel} are
\bes\label{geom}
\begin{align}
& D_\mu D^\mu \vphi + \frac{\vphi}{2|\vphi|}\!\left(\frac{P_{|\vphi|}}{4} F_{\mu\nu}F^{\mu\nu} + V_{|\vphi|} \right)\! =0,\\ \label{meqsc}
& \partial_\mu \left(P F^{\mu\nu}\right) = J^\nu,
\end{align}
\ees
where the current is $J_{\mu} = ie(\overline{\vphi} D_{\mu} \vphi-\vphi\overline{D_{\mu}\vphi})$, $P_{|\vphi|}=dP/d|\vphi|$ and $V_{|\vphi|} = \partial V/\partial{|\vphi|}$. Spacetime translation symmetry yields the following energy-momentum tensor
\be\label{emtgeneral}
T_{\mu\nu} = P F_{\mu\lambda}\tensor{F}{^\lambda_\nu} +\overline{D_{\mu} \vphi}D_{\nu} \vphi + \overline{D_{\nu} \vphi}D_{\mu} \vphi - \eta_{\mu\nu} \LL.
\ee
By setting $\nu=0$ in equation \eqref{meqsc} and considering static configurations, the Gauss' law is satisfied for $A_0=0$, which makes the vortex electrically neutral. In order to search for static and rotationally symmetric solutions, we consider the standard possibility
\be\label{ansatz}
\vphi = g(r)e^{in\theta}, \quad\quad \vec{A} = -{\frac{\hat{\theta}}{er}(a(r)-n)},
\ee
where $n\in\mathbb{Z}$ stands for the vorticity. In this case, the functions $a(r)$ and $g(r)$ obey the boundary conditions
\bes\label{bc}
\bal
g(0) &=0, & a(0)&=n,\\
g(\infty)&=v, & a(\infty)&=0.
\eal
\ees
Here, $v$ is a parameter involved in the symmetry breaking of the potential. Considering the fields described by equations \eqref{ansatz}, the magnetic field takes the form
\be\label{b}
B = -F^{12} = -\frac{a^\prime}{er},
\ee
where the prime represents the derivative with respect to the radial coordinate $r$. By using this, one can show that the magnetic flux can be integrated to give
\be\label{mflux}
\Phi =2\pi\int_0^\infty rdr B = \frac{2\pi n}{e},
\ee
so it is quantized. The equations of motion \eqref{geom} with the static fields \eqref{ansatz} become
\bes\label{secansatz}
\begin{align}
\frac{1}{r} \left(r g^\prime\right)^\prime &= \frac{a^2g}{r^2} + \frac{P_g{a^\prime}^2}{4e^2r^2} + \frac12 V_g, \\
r\left(P\frac{a^\prime}{er}\right)^\prime &= 2eag^2.
\end{align}
\ees
The nonvanishing components of the energy-momentum tensor \eqref{emtgeneral} with the ansatz \eqref{ansatz} are
\bes
\bal\label{rhoans}
T_{00} &= P(g) \frac{{a^\prime}^2}{2e^2r^2} + {g^\prime}^2 +\frac{a^2g^2}{r^2} + V(g), \\
T_{12} &= \left({g^\prime}^2 - \frac{a^2g^2}{r^2}\right) \sin(2\theta), \\ 
T_{11} &= P(g) \frac{{a^\prime}^2}{2e^2r^2} + {g^\prime}^2\left(2\cos^2\theta-1\right) \nonumber\\
       &\hspace{4mm} + \frac{a^2g^2}{r^2}\left(2\sin^2\theta -1\right) -V(g), \\ 
T_{22} &= P(g) \frac{{a^\prime}^2}{2e^2r^2} + {g^\prime}^2\left(2\sin^2\theta-1\right) \nonumber\\
       &\hspace{4mm} + \frac{a^2g^2}{r^2}\left(2\cos^2\theta -1\right) -V(g).
\eal
\ees
where $\rho\equiv T_{00}$ defines the energy density and $T_{ij}$ stands for the stress tensor. The equations of motion \eqref{secansatz} are of second order with couplings between $a(r)$ and $g(r)$, which make them hard to solve. In order to simplify the problem, it is interesting to reduce the problem to the solution of first order equations. To do so, we use the Bogomol'nyi procedure and rewrite the energy density \eqref{rhoans} as
\be
\begin{aligned}
	\rho &= \frac{P(g)}{2}\left(\frac{a^\prime}{er} \pm \frac{e(v^2-g^2)}{P(g)}\right)^2 +  \left(g^\prime \mp \frac{ag}{r} \right)^2 \\
	     &\hspace{4mm} + V -\frac{e^2}{2}\frac{\left(v^2-g^2\right)^2}{P(g)}  \mp \frac{1}{r} \left(a\left(v^2-g^2\right)\right)^\prime.
\end{aligned}
\ee
If the potential is chosen to be
\be\label{potgen}
V(|\vphi|) =  \frac{e^2}{2}\frac{\left(v^2-|\vphi|^2\right)^2}{P(|\vphi|)},
\ee
the energy becomes
\be
\begin{aligned}\label{ene}
	E &= 2\pi\int_0^\infty r\,dr\,\frac{P(g)}{2}\left(\frac{a^\prime}{er}\pm  \frac{e(v^2-g^2)}{P(g)}\right)^2\\
	&\hspace{4mm}+ 2\pi\int_0^\infty r\,dr\,\left(g^\prime \mp \frac{ag}{r} \right)^2 + E_B,
\end{aligned}
\ee
where 
\be\label{ebogo}
\begin{split}
	E_B &= \mp 2\pi\int_0^\infty dr\,\left(a\left(v^2-g^2\right) \right)^\prime \\
	    &= 2\pi|n|v^2.
\end{split}
\ee
Then, the energy is bounded by $E_B$, i.e., $E\geq E_B$. If the solutions obey the first order equations
\bes\label{fovortex}
\bal\label{fog}
g^\prime &= \pm\frac{ag}{r}, \\ \label{foa}
-\frac{a^\prime}{er} &=\pm \frac{e\left(v^2-g^2\right)}{P(g)},
\eal
\ees
the Bogomol'nyi bound is saturated, with the energy minimized to $E=E_B$. One can show that the pair of equations for the lower signs are related to the one for the upper signs by the change $a\to-a$. Furthermore, the presence of the above first order equations also yields stress-free energy-momentum tensor, $T_{ij}=0$, which leads to solutions stable under rescaling; see Ref.~\cite{godvortex}. They also allow us to write the energy density in the form
\be\label{rhov}
\rho =  P(g) \frac{{a^\prime}^2}{e^2r^2} + 2{g^\prime}^2.
\ee

Before going on and investigate the Chern-Simons-Higgs model, let us study some examples of Maxwell-Higgs systems that are controlled by the generalized magnetic permeability $P(|\vphi|)$. We first note from \eqref{potgen} that the presence of first order equations requires an algebraic constraint between the potential $V(|\vphi|)$ and the magnetic permeability $P(|\vphi|)$. Also, for simplicity, we work with dimensionless fields and set $e=1$ and $v=1$, and also, consider unit vorticity, $n=1$, which appears for the upper signs of the first order equations \eqref{fovortex}.

\subsection{First example}

\begin{figure}[htb]
\centering
\includegraphics[width=5cm]{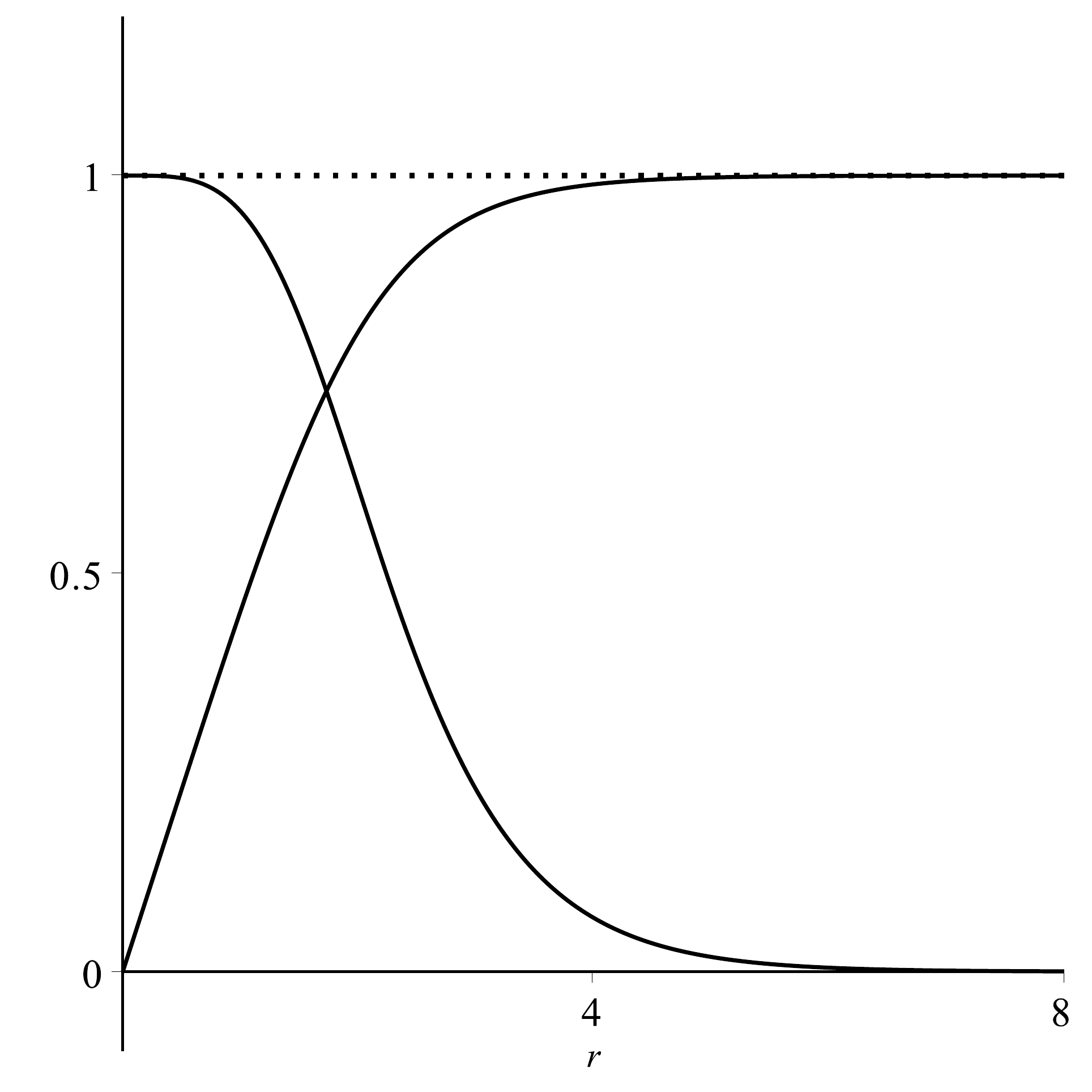}
\caption{The vortex solutions $a(r)$ (descending line) and $g(r)$ (ascending line) of Eqs.~\eqref{fov1}.}
\label{fig1}
\end{figure}

Our first example arises with the magnetic permeability governed by
\be\label{P1}
P(|\vphi|)=\frac{1}{|\vphi|^2}.
\ee
In this case, the potential \eqref{potgen} is a sixth-degree polynomial in the scalar field and the first order equations \eqref{fovortex} with the upper signs become
\bes\label{fov1}
\bal
g^\prime &= \frac{ag}{r}, \\
-\frac{a^\prime}{r} &= g^2\left(1-g^2\right).
\eal
\ees
Due to the choice of $P(|\vphi|)$ in Eq.~\eqref{P1}, we see that the form of the above first order equations makes $a^\prime$ vanish at $r=0$, where $g=0$; see the boundary conditions in Eq.~\eqref{bc}. We then expect to have the function $a(r)$ presenting a plateau near the origin. In this sense, the singular behavior of $P(|\vphi|)$ as $\vphi$ tends to zero is compensated by the vanishing of $a^\prime$, which then nullifies the magnetic field as one approaches the center of the structure. Even though the above equations are of first order, we have been unable to find analytical solutions, so we conduct the investigation numerically and display the solutions in Fig.~\ref{fig1}.

\begin{figure}[htb]
\centering
\includegraphics[width=4.2cm]{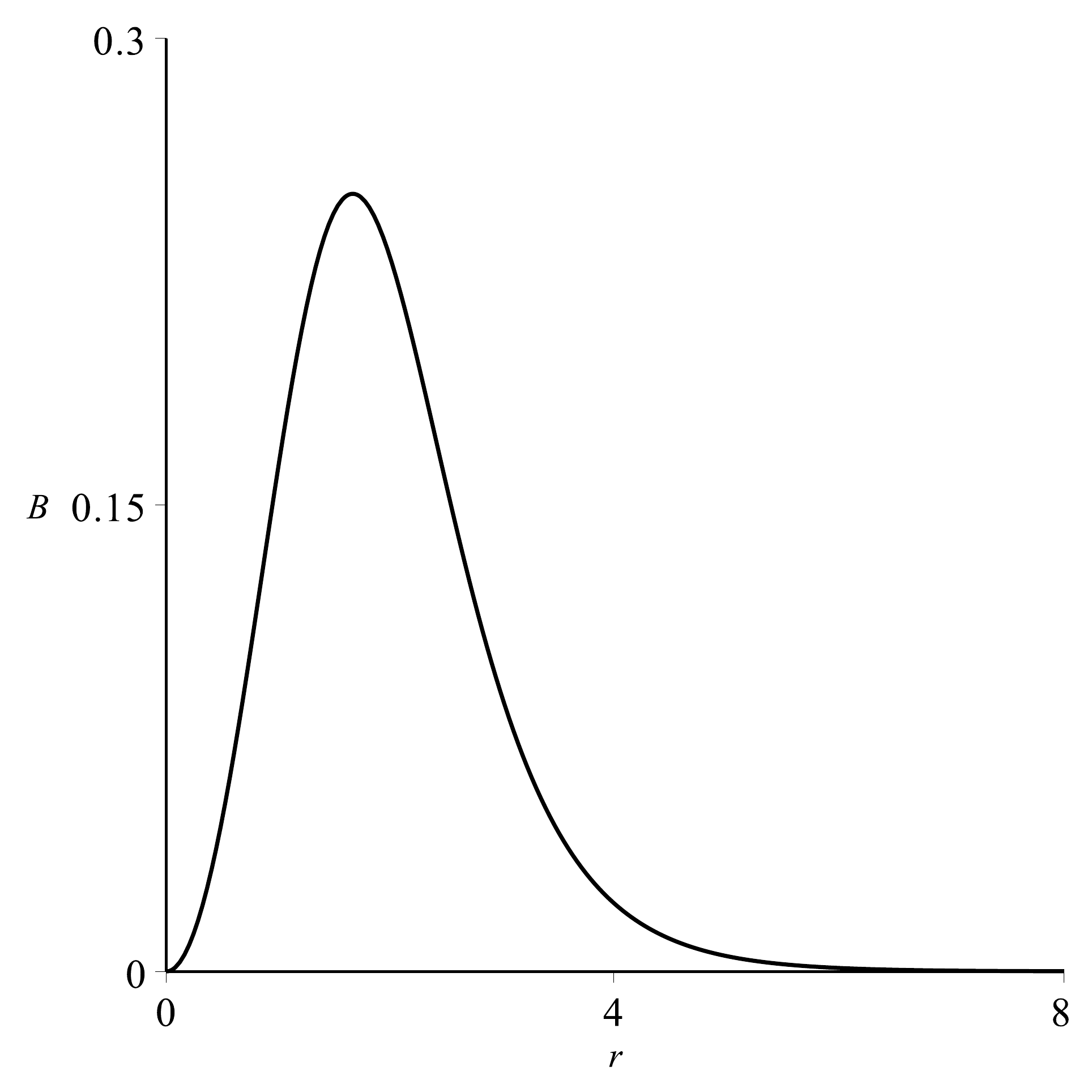}
\includegraphics[width=4.2cm]{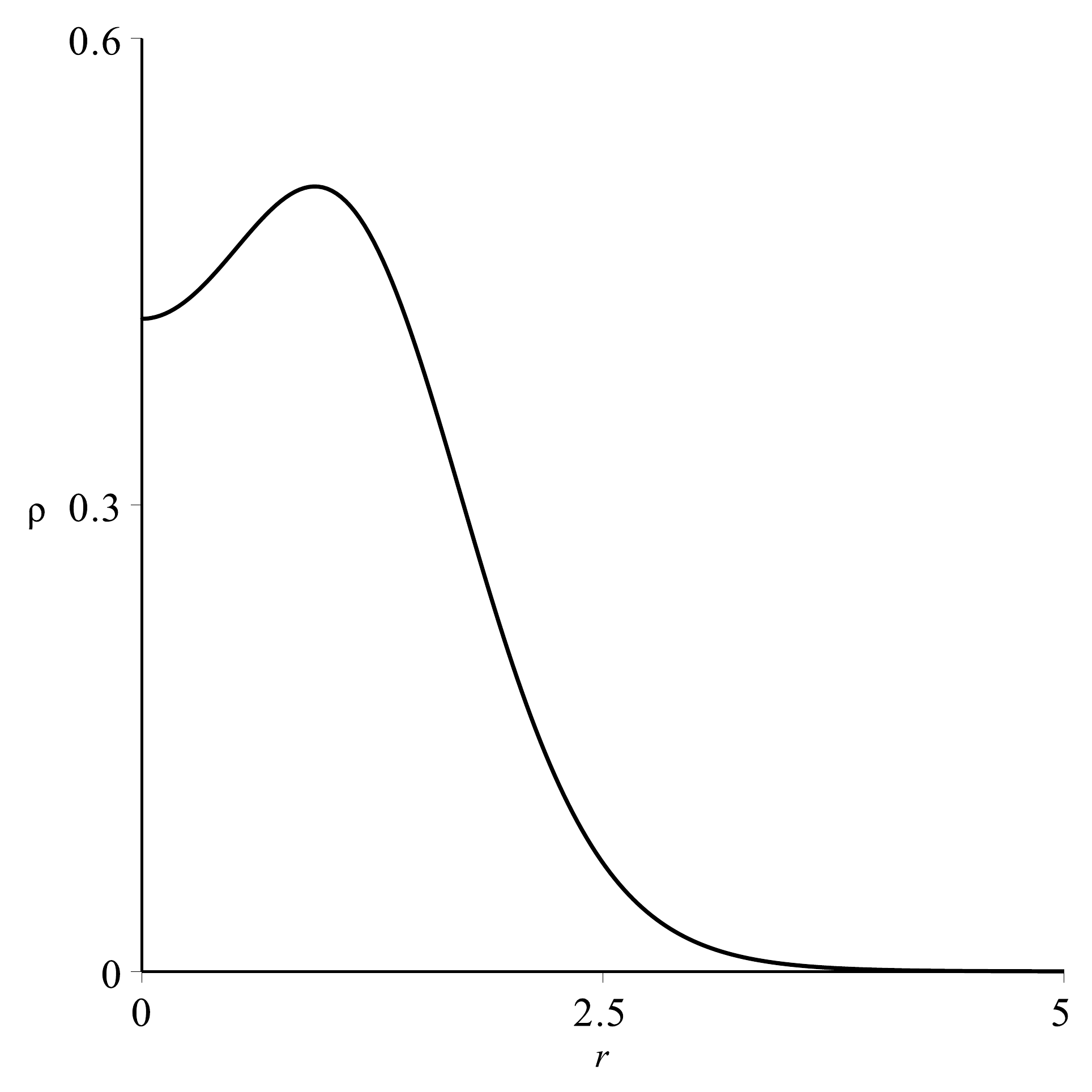}
\caption{The magnetic field (left) and the energy density (right) of the vortex described by the solutions of Eqs.~\eqref{fov1}.}
\label{fig2}
\end{figure}

The magnetic field and the energy density can be calculated through Eqs.~\eqref{b} and \eqref{rhov}. They are shown in Fig.~\ref{fig2}. We see that both of these quantities possess a valley at the origin, though in qualitatively different ways. This is the effect of the vanishing behavior of $a^\prime$ at the origin. By integrating the energy density of the vortex in the plane, we find $E=2\pi$, which matches with Eq.~\eqref{ebogo} for unit $n$ and $v$.

\begin{figure}[t]
\centering
\includegraphics[width=4.2cm]{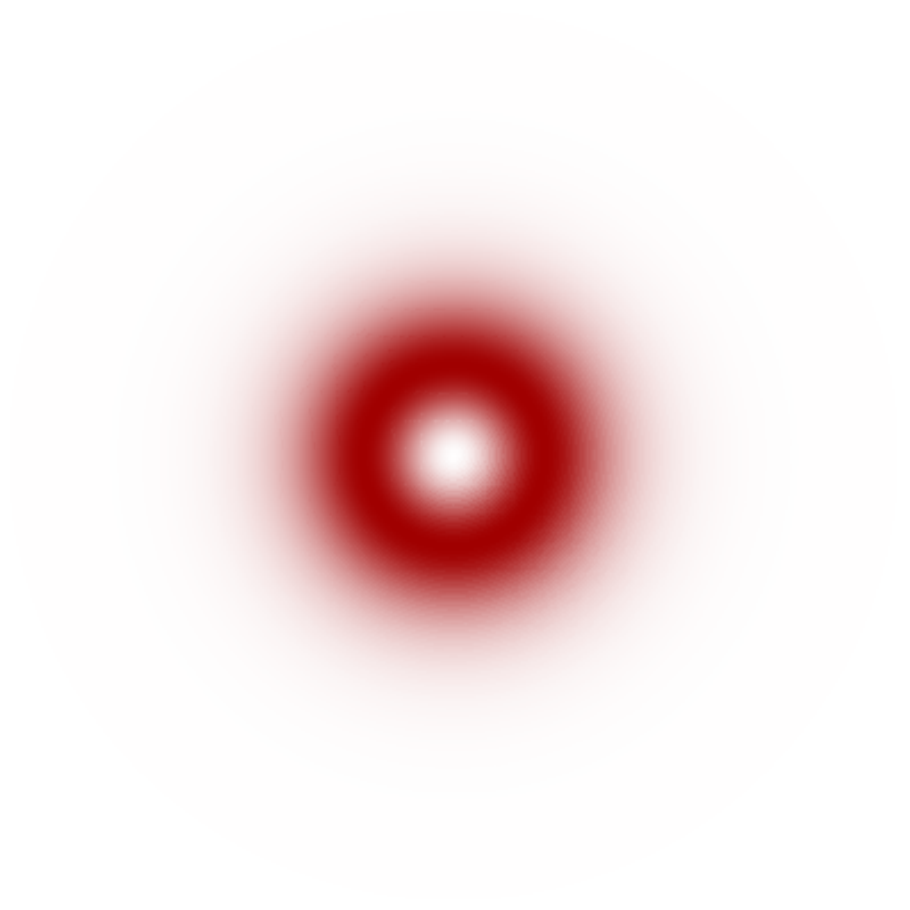}
\includegraphics[width=4.2cm]{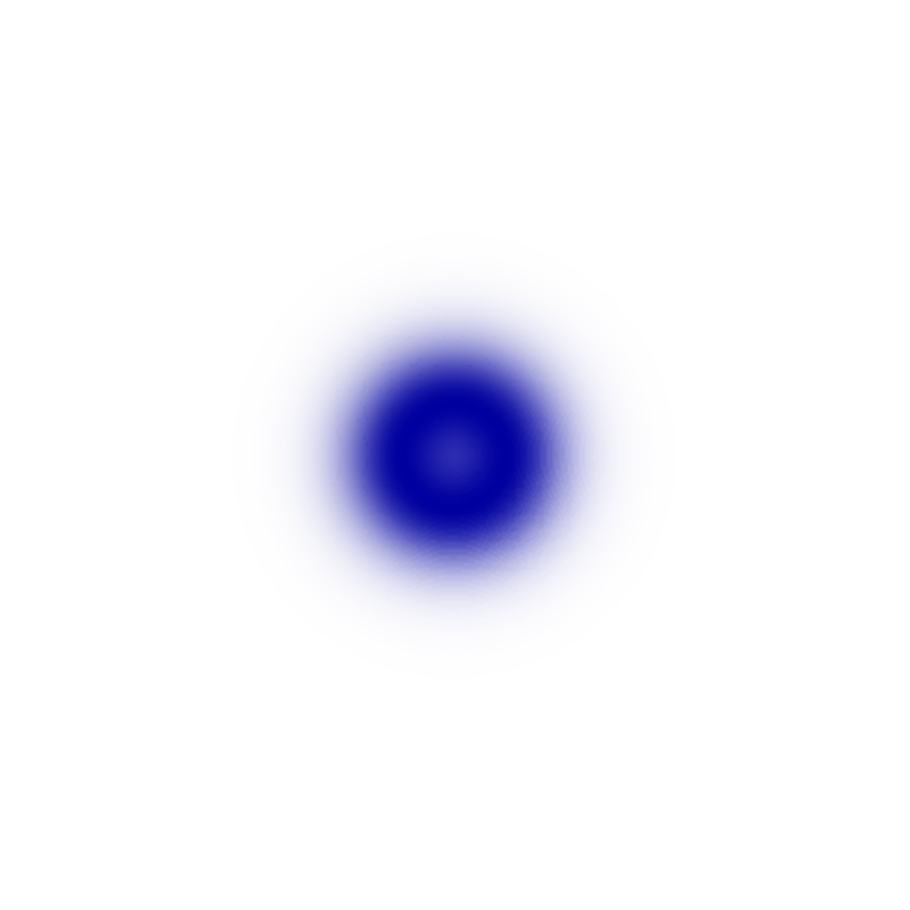}
\caption{The magnetic field (left) and the energy density (right) of the vortex in the plane, associated to the solutions of Eqs.~\eqref{fov1} for $r\in[0,8]$. The darkness of the color is related to the intensity of the corresponding quantities.}
\label{fig3}
\end{figure}

The presence of the valley in the vortex core motivated us to depict the magnetic field and the energy density in the plane; they are shown in Fig.~\ref{fig3}. It is evident that the magnetic field vanishes at the origin, despite the presence of some matter there, since the energy density does not vanish at $r=0$.

\subsection{Second example}
We consider another case, engendered by
\be\label{P2}
P(|\vphi|) = \frac{1}{|\vphi|^2\left(\alpha-|\vphi|^2\right)^2},
\ee
with $\alpha\in(0,1)$. This choice diverges at $|\vphi|\to0$ and at $|\vphi|\to\sqrt{\alpha}$, so it forces $a^\prime$ to vanish at the same values, leaving the magnetic field with a richer internal structure. It also makes the potential \eqref{potgen} to have a generalized form, up to the tenth degree in the scalar field. This polynomial generalization can be thought as coming from higher order operators contributing to the effective theory, inducing higher order (structure) effects in a quantum theory of vortices. We also recall that in the holographic model studied in \cite{HO2}, the potential involved hyperbolic functions of the scalar field. As we are going to show, the choice \eqref{P2} will lead to a magnetic field with a concentric double-ring structure.

\begin{figure}[t]
\centering
\includegraphics[width=5cm]{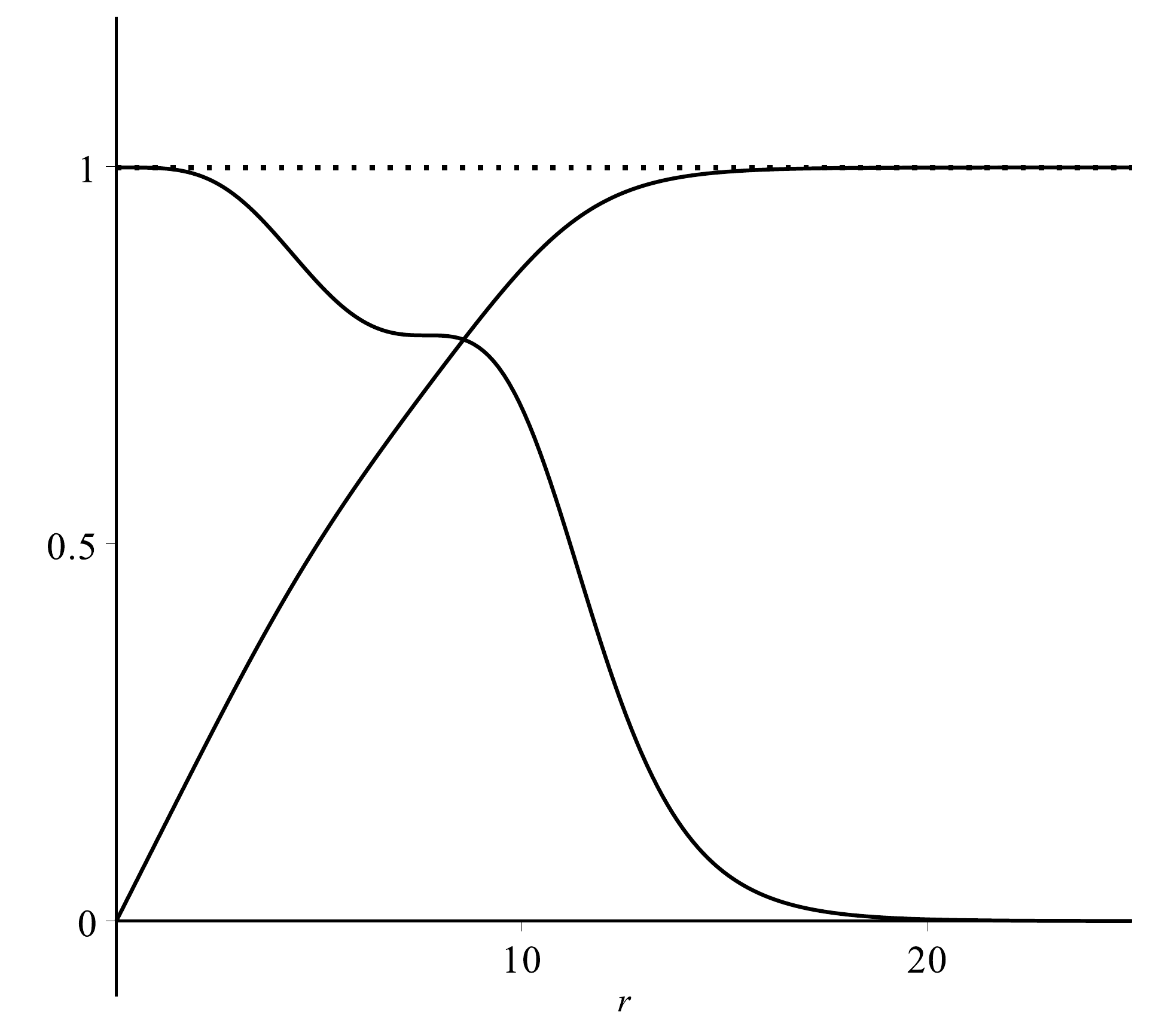}
\caption{The vortex solutions $a(r)$ (descending line) and $g(r)$ (ascending line) of Eqs.~\eqref{fov2} with $\alpha=0.5$.}
\label{fig4}
\end{figure}

In this case, first order equations \eqref{fovortex} with the upper signs become
\bes\label{fov2}
\bal
g^\prime &= \frac{ag}{r}, \\
-\frac{a^\prime}{r} &= g^2\left(\alpha-g^2\right)^2\left(1-g^2\right).
\eal
\ees
This model is richer than the previous one, since the function \eqref{P2} makes $a^\prime$ vanish not only at the origin, but also in the point where $g^2=\alpha$, which always exists for $\alpha\in(0,1)$. Thus, we expect the presence of two plateaux in $a(r)$. As before, we must use numerical methods to solve the equations above, and in Fig.~\ref{fig4} we depict the solutions for $\alpha=0.5$.

\begin{figure}[htb]
\centering
\includegraphics[width=4.2cm]{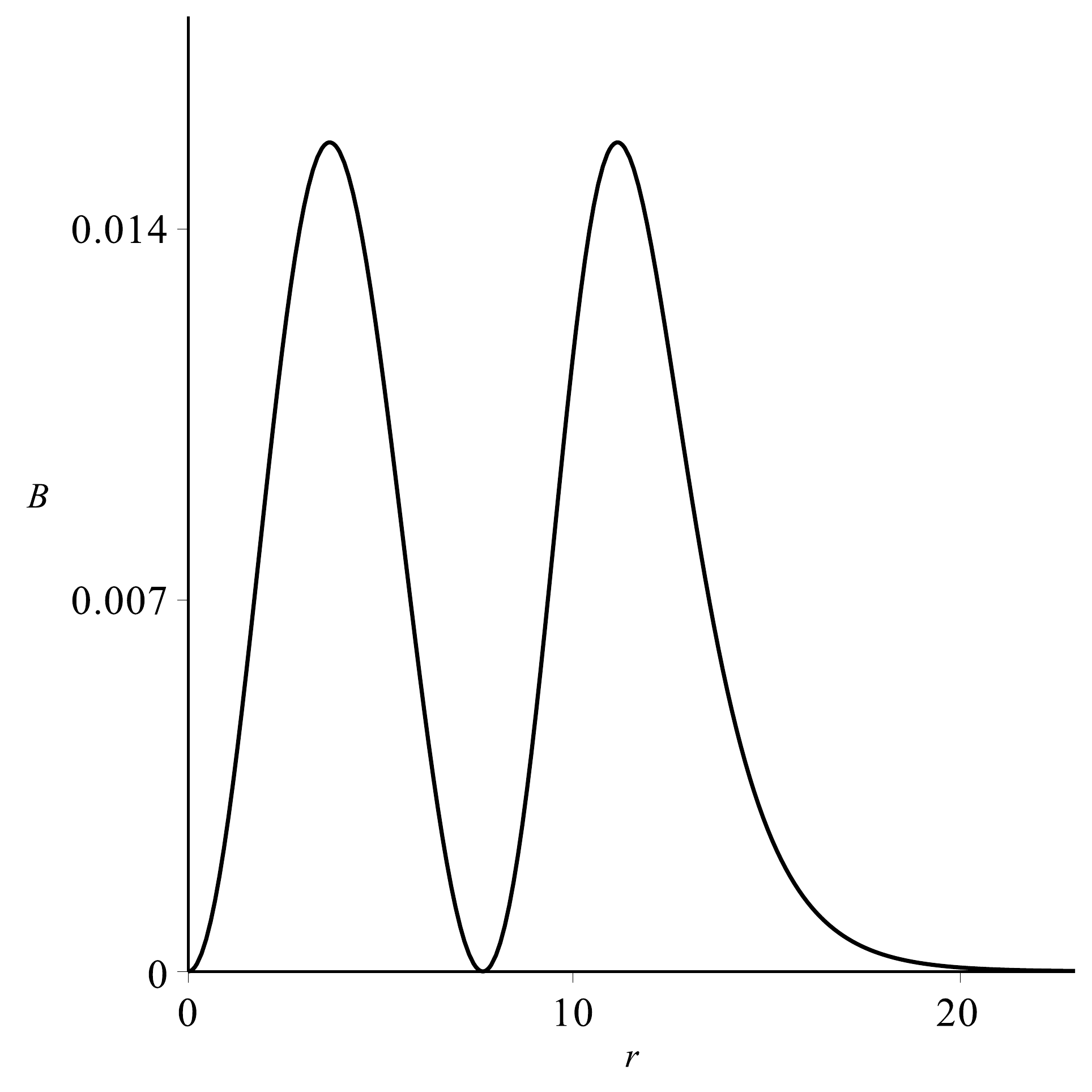}
\includegraphics[width=4.2cm]{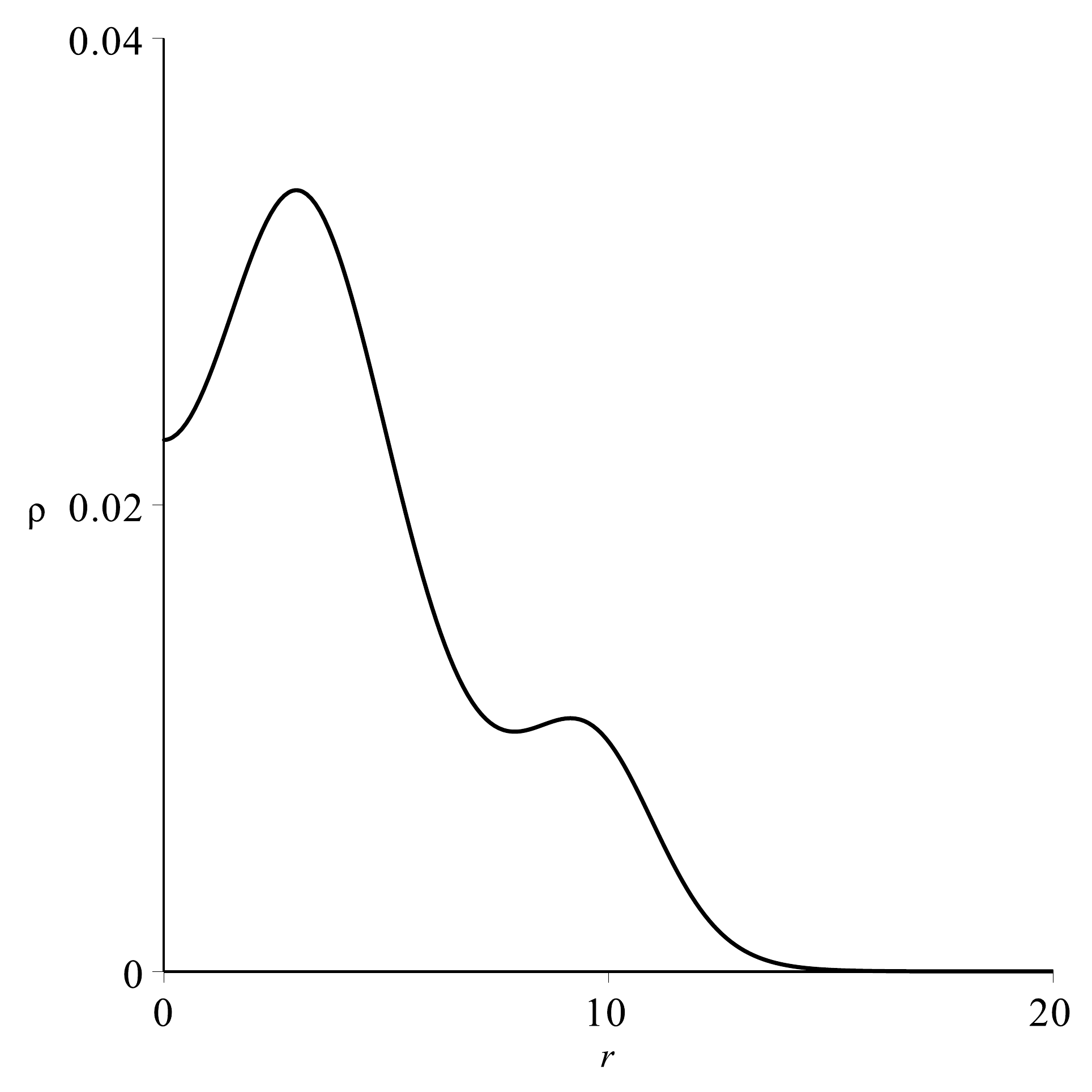}
\caption{The magnetic field (left) and the energy density (right) of the vortex, associated to the solutions of Eqs.~\eqref{fov2} with $\alpha=0.5$.}
\label{fig5}
\end{figure}

The magnetic field and the energy density can be calculated through Eqs.~\eqref{b} and \eqref{rhov}. In Fig.~\ref{fig5}, we show them for the solutions of Eqs.~\eqref{fov2} with $\alpha=0.5$. Due to the behavior of $a(r)$, these quantities have two valleys, one around the origin and the other around the point where $g^2=\alpha$. Notice that this effect is stronger in the magnetic field than in the energy density. By integrating the energy density of the vortex, we find $E=2\pi$, which matches with Eq.~\eqref{ebogo} for unit $n$ and $v$.

\begin{figure}[t]
\centering
\includegraphics[width=4.2cm]{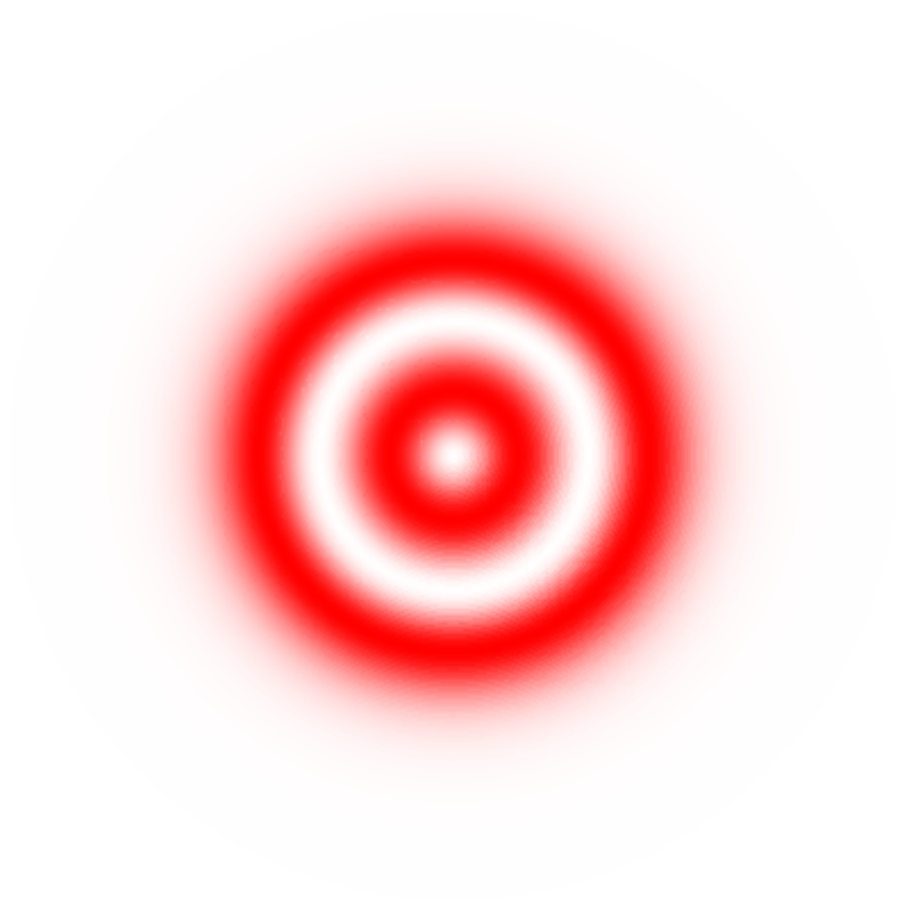}
\includegraphics[width=4.2cm]{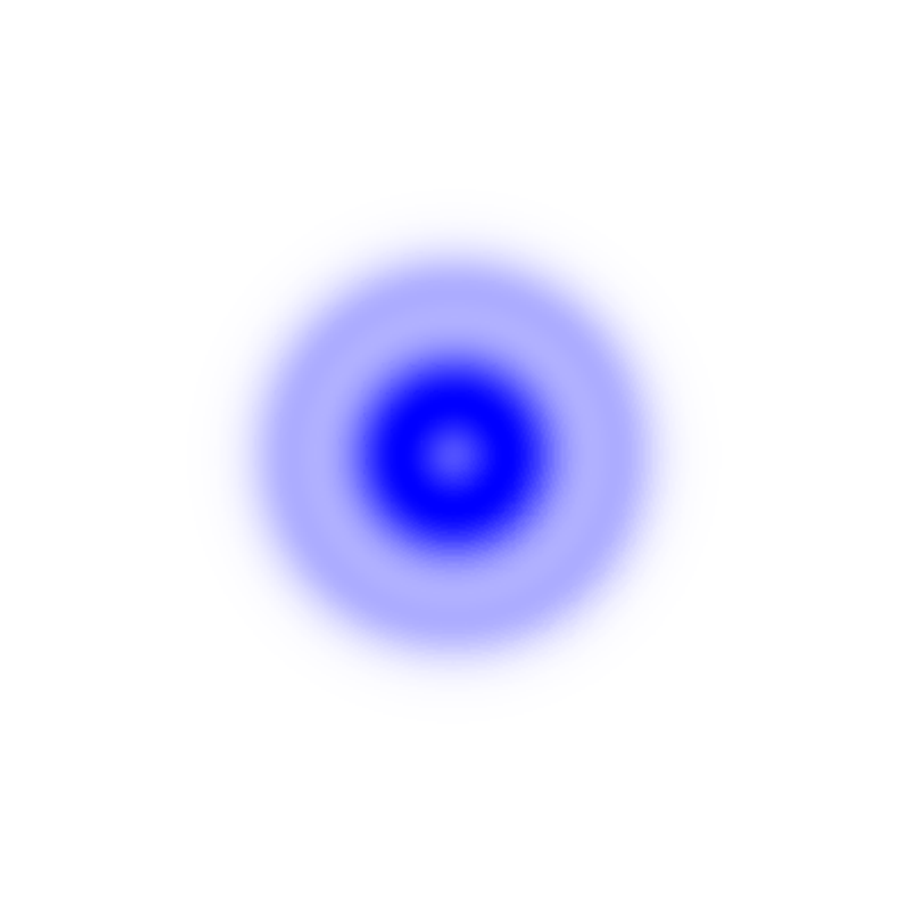}
\caption{The magnetic field (left) and the energy density (right) of the vortex in the plane, associated to the solutions of Eqs.~\eqref{fov2} for $\alpha=0.5$, with $r\in[0,25]$. The darkness of the color is related to the intensity of the corresponding quantities.}
\label{fig6}
\end{figure}

The behavior of $a(r)$ displayed in Fig.~\ref{fig5} shows that the vortex engenders internal structure. This feature motivated us to depict the magnetic field and the energy density in the plane in Fig.~\ref{fig6}, which shows the presence of two ringlike substructures, much more accentuated in the magnetic field, as indicated by the behavior presented in Fig.~\ref{fig5}.

\section{Chern-Simons-Higgs Models}\label{cs}

We consider the same spacetime features of the previous section, but now we substitute the Maxwell term by the Chern-Simons one
\begin{equation}\label{lcs}
\LL = \frac{\kappa}{4}\epsilon^{\alpha\beta\gamma}A_\alpha F_{\beta\gamma} + K(|\vphi|)\overline{D_\mu \vphi}D^\mu\vphi -V(|\vphi|).
\end{equation}
In the above expression, the fields are denoted as in the previous setup. Here, however, the dynamical term of the gauge field, that is, the Chern-Simons term, does not admit the factor $P(|\vphi|)$ as we did in the previous model, because it would break gauge invariance. In this sense, $\kappa$ has to be a real constant. In order to generalize the model, however, we introduced an arbitrary nonnegative function $K(|\varphi|)$ in the kinetic term of the scalar field, which effectively modifies the coupling between the scalar and gauge fields. This is also a standard modification of the complex scalar field Lagrangian, especially in supersymmetric theories. Function $K(|\vphi|)$ is introduced as a second derivative of the K\"ahler potential, which also defines a metric on the space of scalar fields. This kind of interaction appeared before, for instance, in \cite{g7}, to describe Bogomol'nyi equations in a generalized Maxwell-Higgs model, and also in Ref.~\cite{HO1}, in which a holographic model of a strongly interacting superfluid is considered in the context of a Maxwell field and a charged complex scalar field that are coupled to gravity.

We go on and consider $A^\mu = (A^0,\vec{A})$, in a manner that the electric and magnetic fields are now given by
\be\label{eb}
E^i = F^{i0} = -\dot{A}^i - \partial_i A^0 \quad\text{and}\quad B = -F^{12},
\ee
with $(E_x,E_y)\equiv (E^1,E^2)$. One may vary the action associated to the Lagrangian in Eq.~\eqref{lcs} to get the equations of motion for the scalar and gauge fields
\bes\label{gcseom}
\begin{align}
 D_\mu (K D^\mu\vphi)&= \frac{\vphi}{2|\vphi|}\left(K_{|\vphi|}\overline{D_\mu \vphi}D^\mu\vphi -V_{|\vphi|} \right), \\ \label{cseqs}
 \frac{\kappa}{2} \epsilon^{\lambda\mu\nu}F_{\mu\nu} &= J^\lambda,
\end{align}
\ees
where $J_\mu = ieK(|\vphi|)(\bar{\vphi}D_\mu \vphi-\vphi\overline{D_\mu\vphi})$. Spacetime translation symmetry gives the energy-momentum tensor
\be\label{tmunu}
T_{\mu\nu}=K(|\vphi|)\left( \overline{D_\mu \vphi}D_\nu \vphi + \overline{D_\nu \vphi}D_\mu \vphi\right) - \eta_{\mu\nu}\LL.
\ee

In order to investigate the presence of vortexlike solutions, we consider the same $\vphi$ and $\vec {A}$ of Eq.~\eqref{ansatz}, and add  $A_0=A_0(r)$.  Therefore, the magnetic field is given as in Eq.~\eqref{b}, which leads to the quantized flux \eqref{mflux}. Since $A_0\neq0$, the vortex possesses a nonvanishing electric field
\be
E^i = -\frac{x^i}{r} A_0^\prime.
\ee
Due to this, the vortex carries electric charge
\be
\begin{split}
Q = 2\pi\int rdr J^0= -\kappa\Phi.
\end{split}
\ee
Therefore, the electric charge is proportional to the magnetic flux and so also quantized. In the Chern-Simons model, the equations of motion \eqref{gcseom} can be written in the form
\bes\label{eomcsansatz}
\begin{align}
\frac{1}{r} \left(rK g^\prime\right)^\prime + K g \left(e^2 A_0^2-\frac{a^2}{r^2} \right)+& \nonumber\\
+ \frac12 \left(\left(e^2g^2A_0^2-{g^\prime}^2-\frac{a^2g^2}{r^2}\right)K_{|\vphi|} -V_{|\vphi|} \right) &= 0, \\ \label{a0csansatz}
 \frac{a^\prime}{r} + \frac{2K e^3g^2 A_0}{\kappa} &= 0, \\
 {A_0^\prime} + \frac{2K ea g^2}{\kappa r} &= 0.
\end{align}
\ees
The energy density is given by the $T_{00}$ component in Eq.~\eqref{tmunu}. Combining it with Eq.~\eqref{a0csansatz}, we get
\be\label{rhoansC}
\rho =\frac{\kappa^2}{4e^4} \frac{{a^\prime}^2}{r^2 g^2K(g)} + \left({g^\prime}^2+\frac{a^2g^2}{r^2}\right)K(g) + V(g).
\ee
As in the previous section, we rewrite the above energy density in the form
\be
\begin{aligned}
	\rho &=\left(\frac{\kappa a^\prime}{2e^2r\sqrt{g^2K(g)}} \pm \sqrt{V(g)}\right)^2 + K(g)\left(g^\prime\mp\frac{ag}{r}\right)^2 \\
	     &\hspace{4mm}\pm\frac{ag^\prime}{r}\left(\frac{\kappa}{e^2} \frac{d}{dg}\sqrt{\frac{V(g)}{g^2K(g)}} + 2gK(g)\right) \\
	     &\hspace{4mm}\mp\frac{1}{r}\left(\frac{\kappa}{e^2}\,a\,\sqrt{\frac{V(g)}{g^2K(g)}}\right)^\prime.
\end{aligned}
\ee
Thus, if the constraint
\be\label{constraintVK}
\frac{d}{dg} \left(\sqrt{\frac{V}{g^2K}}\,\right) = -\frac{2e^2}{\kappa}gK
\ee
is satisfied, we get the energy
\be
\begin{aligned}
	E &=2\pi\int_0^\infty rdr\left(\frac{\kappa a^\prime}{2e^2r\sqrt{g^2K(g)}} \pm \sqrt{V(g)}\right)^2 \\
	  &\hspace{4mm} + 2\pi\int_0^\infty rdr K(g)\left(g^\prime\mp\frac{ag}{r}\right)^2 + E_B,
\end{aligned}
\ee
where
\be\label{ebcs}
\begin{split}
	E_B &= \mp2\pi\int_0^\infty dr \left(\frac{\kappa}{e^2}\,a\,\sqrt{\frac{V(g)}{g^2K(g)}}\right)^\prime \\
	    &= \frac{2\pi|n|\kappa}{e^2}\lim_{g\to0} \sqrt{\frac{V(g)}{g^2K(g)}}.
\end{split}
\ee
From the above equation, we see that the energy is bounded, i.e., $E\geq E_B$. Thus, if the constraint \eqref{constraintVK} is satisfied and the first order equations
\bes\label{focs}
\bal
g^\prime &= \pm\frac{ag}{r},\\
-\frac{\kappa a^\prime}{2er} &=\pm g\sqrt{KV}, 
\eal
\ees
are obeyed, the energy is minimized to $E=E_B$ as in Eq.~\eqref{ebcs}. 

\begin{figure}[t]
\centering
\includegraphics[width=5cm]{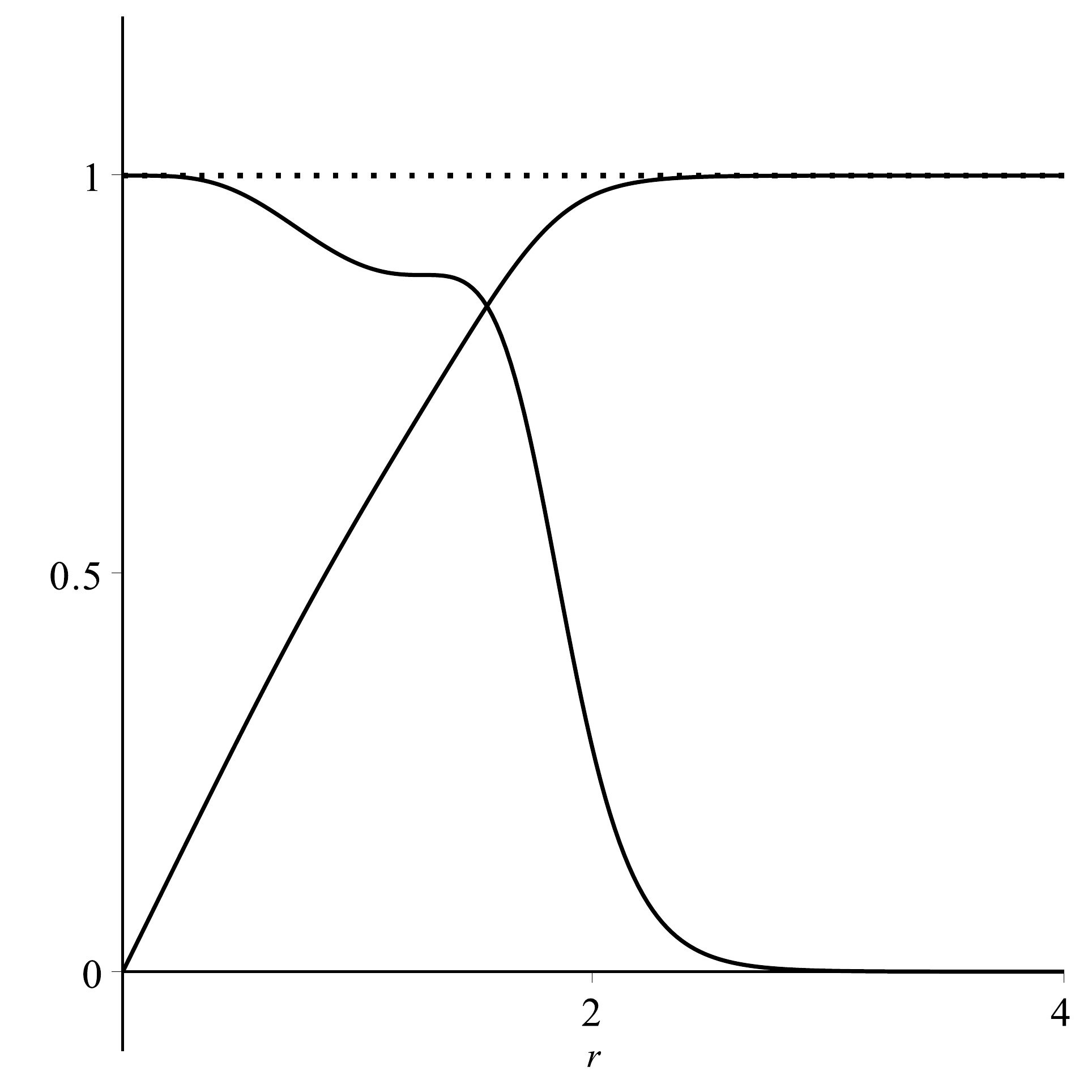}
\caption{The vortex solutions $a(r)$ (descending line) and $g(r)$ (ascending line) of Eqs.~\eqref{focsex}.}
\label{fig7}
\end{figure}

Here we stress that the constraint \eqref{constraintVK} must be solved in a manner that the potential admits a minimum at $|\vphi|=v$, in order to get solutions that obey the boundary conditions \eqref{bc}. Furthermore, one can show that the above first order equations are compatible with the equations of motion \eqref{eomcsansatz}. An interesting feature is that their solutions imply stress-free energy-momentum tensor, $T_{ij}=0$, which is related to stability under rescaling; see Ref.~\cite{godvortex}. By using them, we can write the energy density as
\be\label{rhocs}
\rho =\frac{\kappa^2}{2e^4} \frac{{a^\prime}^2}{r^2 g^2K(g)} + 2K(g){g^\prime}^2.
\ee

The presence of the differential equation \eqref{constraintVK} as the constraint linking $K(|\vphi|)$ and $V(|\vphi|)$ makes the problem harder than in the Maxwell-Higgs system, where $P(|\vphi|)$ and $V(|\vphi|)$ are linked via the algebraic constraint giving by  \eqref{potgen}, so we have to proceed carefully in the current context. As before, for simplicity we also consider dimensionless units and set $e,\kappa,v=1$, and work with unit vorticity, $n=1$, which requires the upper signs in the first order equations \eqref{focs}. Let us consider the pair of functions
\bal
K(|\vphi|) &= 3\left(1-2|\vphi|^2\right)^2,\\
V(|\vphi|) &= 3|\vphi|^2\!\!\left(1-|\vphi|^2\right)^2\!\!\left(1-2|\vphi|^2\right)^2\!\!\left(1-2|\vphi|^2+4|\vphi|^4\right)^2\!\!.
\eal
We note that they automatically satisfy the constraint \eqref{constraintVK}, so the first order equations \eqref{focs} take the form
\bes\label{focsex}
\bal
g^\prime &= \frac{ag}{r},\\
a^\prime &=-6r g^2 \left(1-g^2\right)\left(1-2g^2\right)^2\left(1-2g^2+4g^4\right).
\eal
\ees
We see from the above equations that $a^\prime$ vanishes at $r=0$ (where $g=0$) and at the point where $g^2=1/2$. This behavior suggests the presence of two distinct plateaux around these two points. Unfortunately, we have been unable to find analytical solutions of the above equations, so we use numerical methods and depict the vortex solutions in Fig.~\ref{fig7}, where we clearly note the two plateaux for $a(r)$.

\begin{figure}[t]
\centering
\includegraphics[width=4cm]{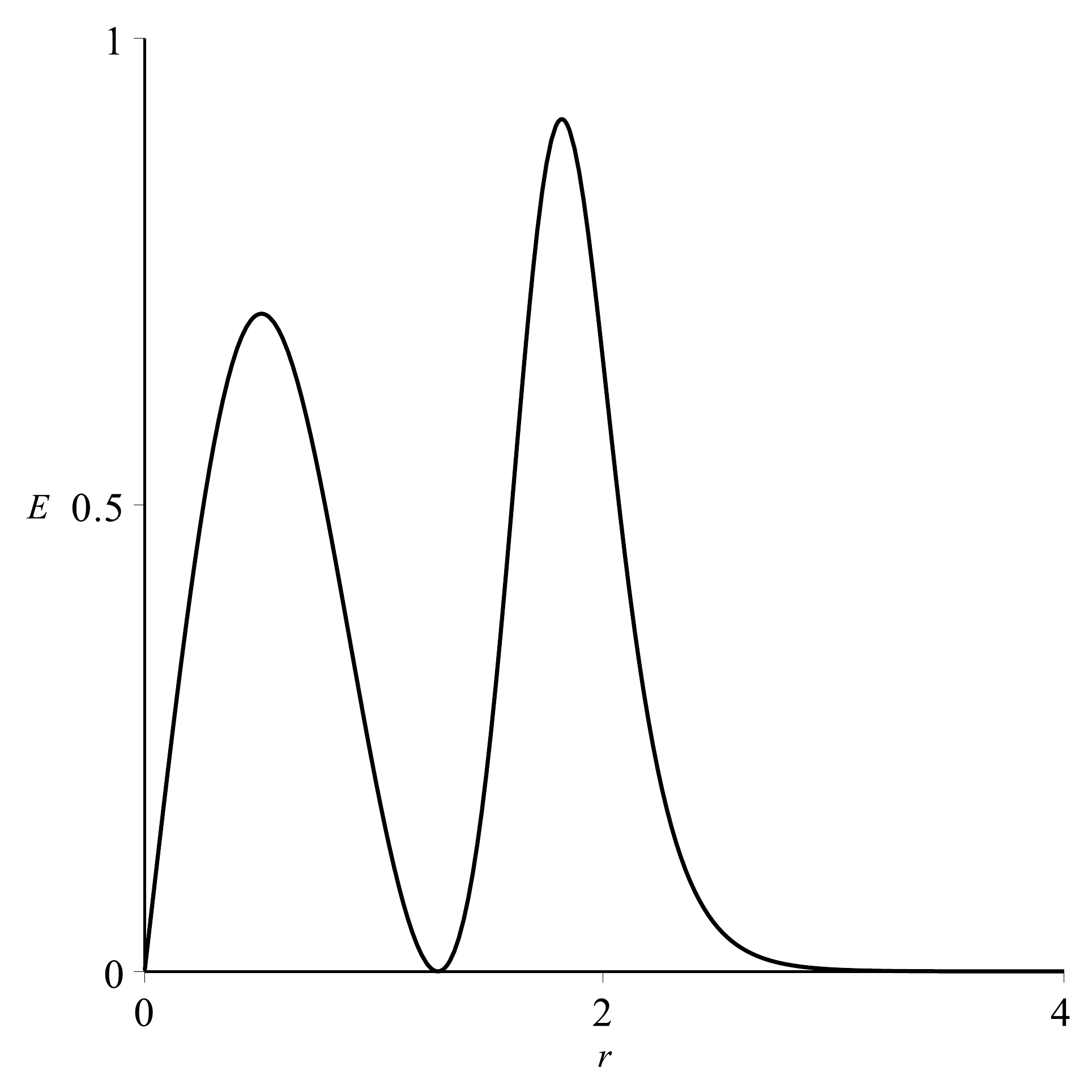}
\includegraphics[width=4cm]{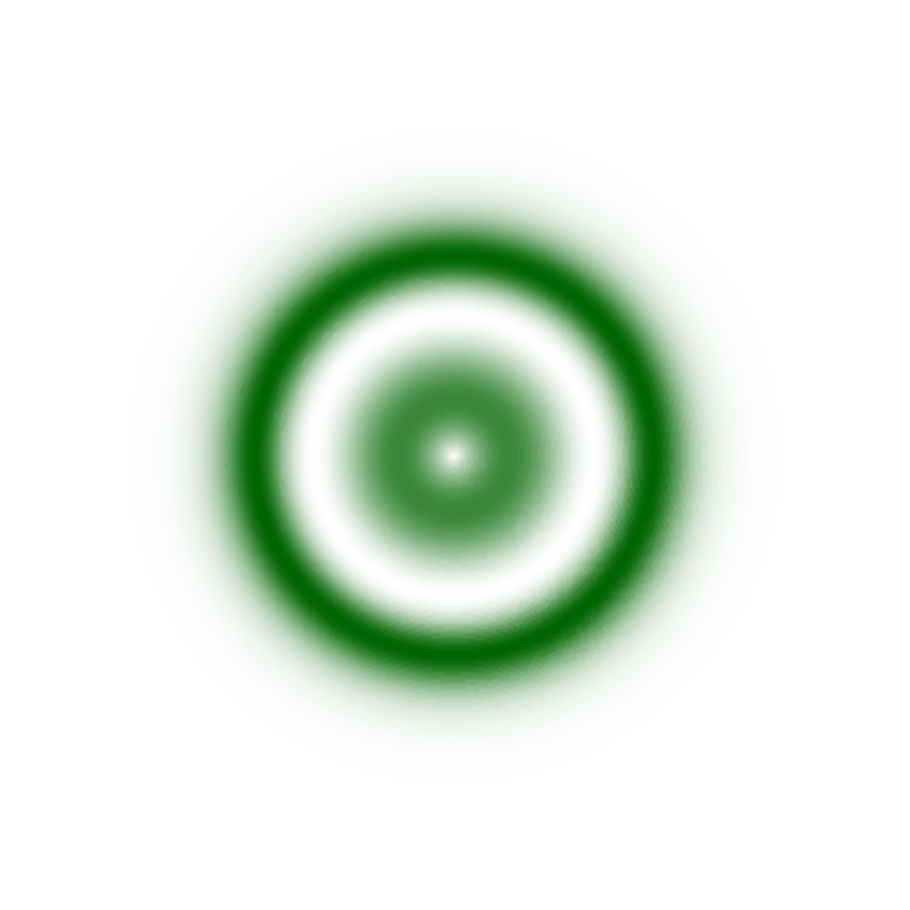}
\includegraphics[width=4cm]{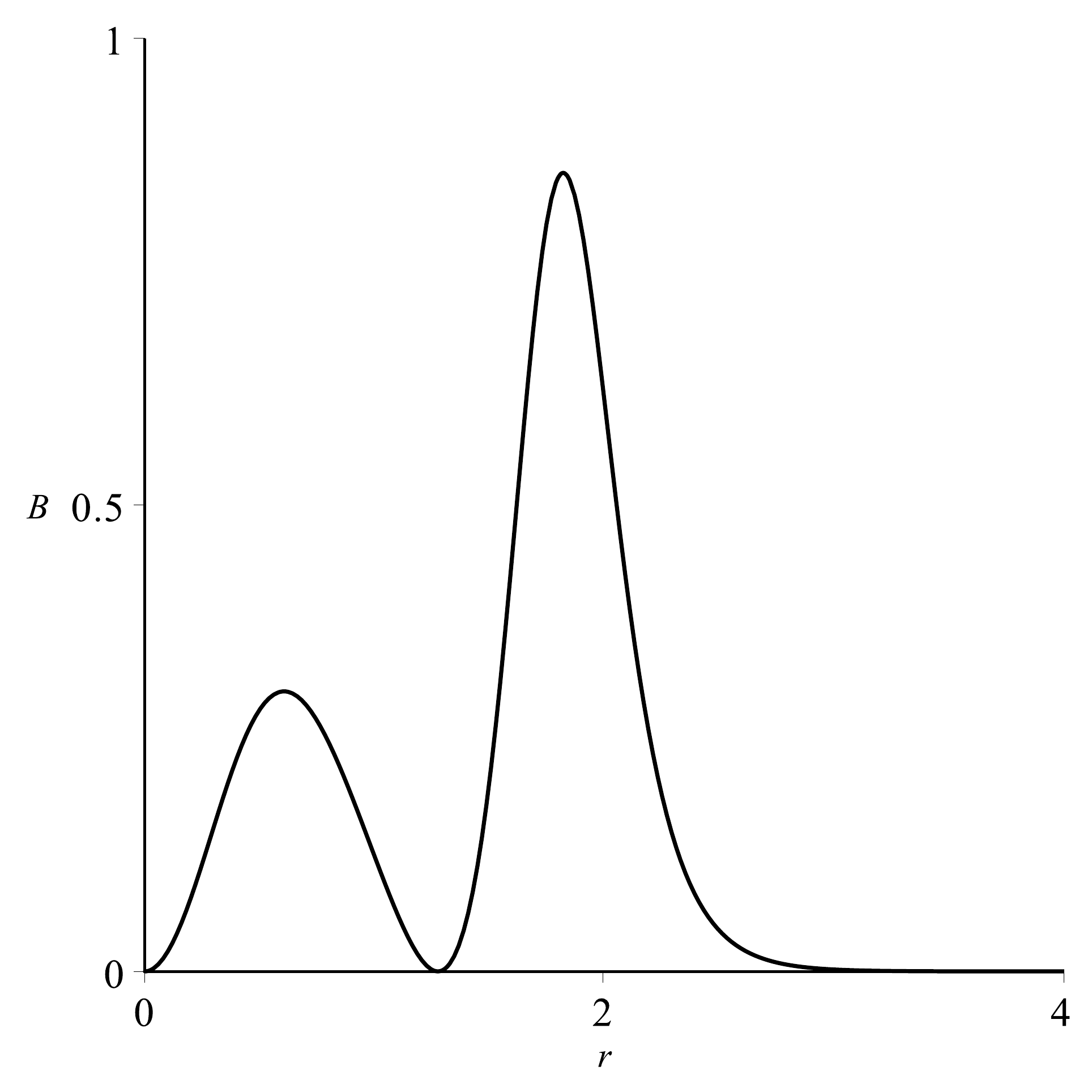}
\includegraphics[width=4cm]{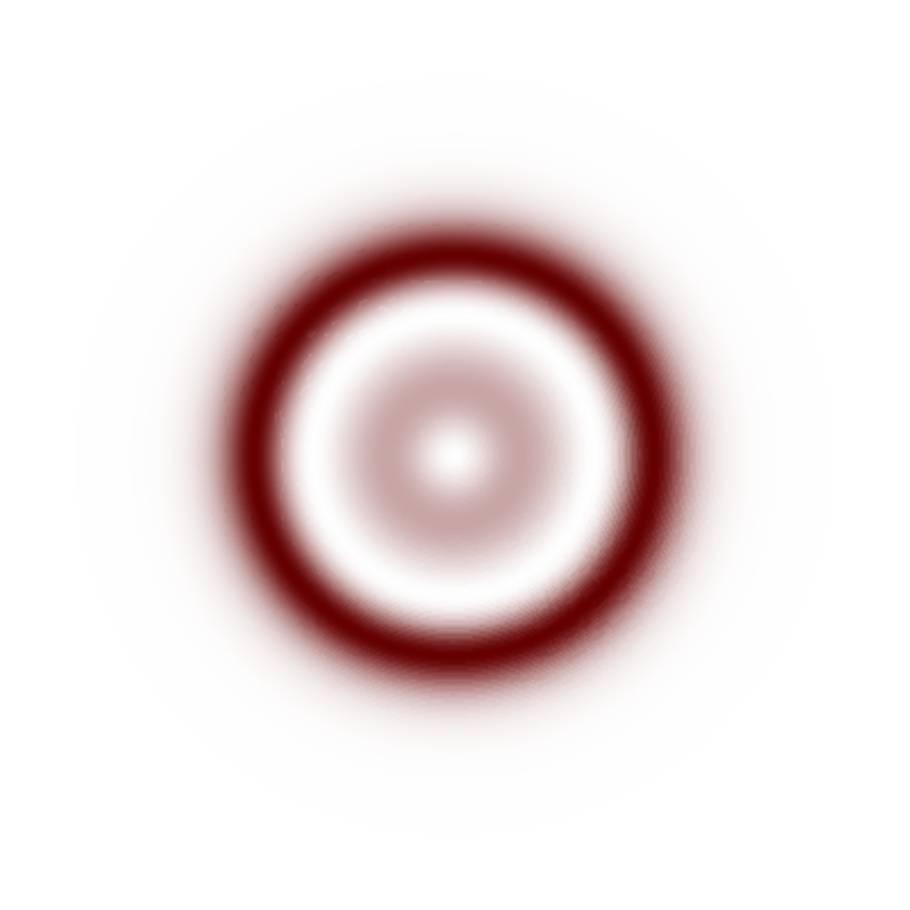}
\includegraphics[width=4cm]{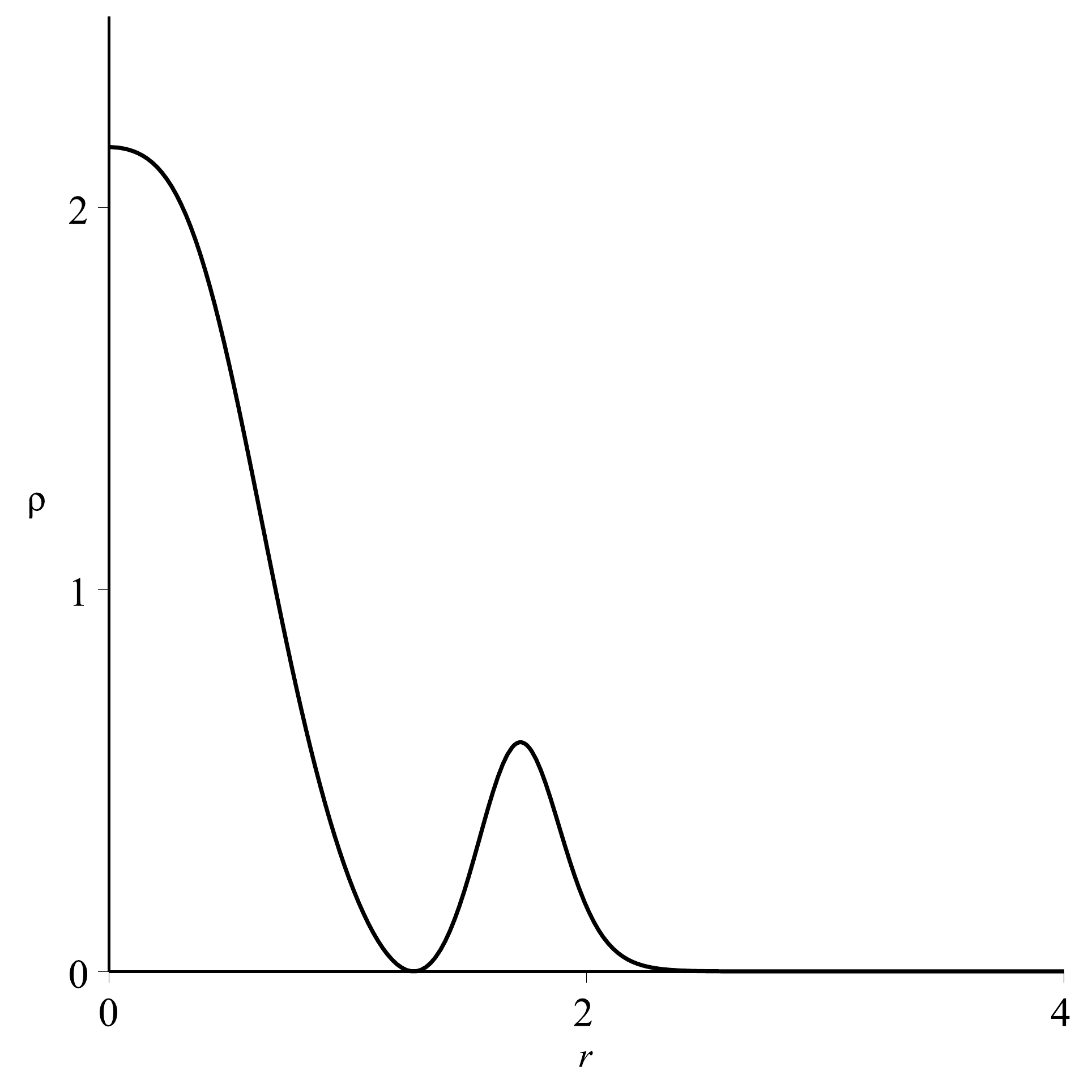}
\includegraphics[width=4cm]{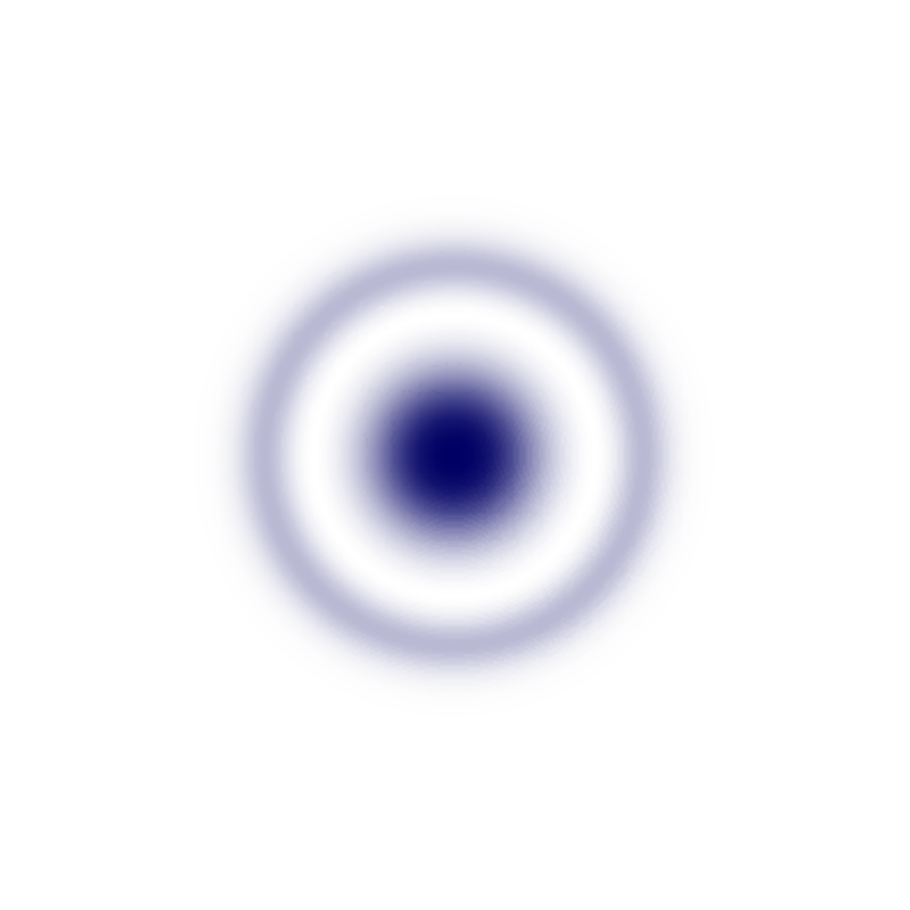}
\caption{The electric field (top), magnetic field (middle) and the energy density (bottom) of the vortex associated to the solutions of Eqs.~\eqref{focsex}. The plots in the right side represent these quantities in the plane for $r\in[0,4]$.}
\label{fig8}
\end{figure} 

By using the numerical solutions, we plot the electric field modulus, magnetic field and energy density in Fig.~\ref{fig8}, including their behavior in the plane. By a numerical integration, we get $E=2\pi$, which is compatible with Eq.~\eqref{ebcs} for $e,\kappa,n=1$. An interesting fact is that both the electric and magnetic fields develop similar two-ring structures, even though the energy density is supported on a disk and a ring around it. Thus, in the vortex core, the matter seems to be unable to generate electric and magnetic fields.

\section{Conclusions}\label{conclusions}

In this work we studied the presence of vortices in two distinct generalized models, one with the gauge dynamics controlled by the Maxwell term, and the other by the Chern-Simons term. We analyzed the two models under the BPS procedure, and although they have very different dynamical behavior, they support first order differential equations that solve the equations of motion for static and rotationally symmetric field configurations. 

The investigation revealed that the Maxwell-Higgs model is capable of supporting electrically neutral vortices that engender interesting internal structure, with the ringlike profile. In the case of the Chern-Simons model, however, the presence of the electric field adds a constraint that restricts the potential and makes the system harder to investigate. In spite of this, we have been able to describe a vortex configuration that is charged electrically, with the electric and magnetic fields endowing similar ringlike structures. The same happens with the energy density, although it does not vanish at the core of the vortex configuration.

The investigation was done with the two distinct models having the same $U(1)$ symmetry. So, differently from the investigation done before in \cite{vortexint}, here we have added no other field, keeping the same symmetry of the original model, so we may conclude that the enhancement of the symmetry with the inclusion of extra degrees of freedom is not mandatory for the appearance of topological vortices with internal structure. We noted a similar behavior in the case of magnetic monopoles, as recently investigated in \cite{magmono}, despite the very different environments required to produce vortices and monopoles.

The Maxwell-Higgs models investigated in Sec. II may be extended to the four-dimensional spacetime, turning the corresponding vortex configurations into cosmic strings \cite{vilenkin,manton,weinberg}. The model considered in Sec.~II.A is described by a polynomial potential, which is degree six in the scalar field, and gives rise to a ring-like structure which is similar to the vortex excitations found before, for instance, in a dipolar Bose-Einstein condensate \cite{CM3}, suggesting that that the dipolar interaction in
the condensate may be controlled by an effective magnetic permeability, which can serve as an indication for the cold atom experiments. The importance of the internal structure of vortices is also stressed in the context of superfluidity; see, e.g., Ref.~\cite{dmi}, where it is argued that in the hydrodynamics description of a superfluid, one has at the very least to include their dipole moments. 

The model examined in Sec. II.B, is defined with an additional parameter $\alpha$, which makes the magnetic field to engender a richer profile, in the form of a concentric double-ring structure. This parameter is a result of an irrelevant deformation of the theory, which might become important at high energies. For example, it may modify the way the vortices interact, evolve and contribute for structure formation in the early Universe, changing the standard scenario. In particular, the mass of the scalar field in this case depends on $\alpha$, and this may affect the cosmological phase transition at high temperature. The model considered in Sec. III is of the Chern-Simons type, so it may be of interest to planar magnetic domain, anyon superconductivity and the Hall effect \cite{fradkin,hubert}.

The models investigated in this paper are simpler than the models considered in Ref.~\cite{vortexint}. They are also partly motivated by possible supersymmetric extensions. Generalized couplings, like the ones introduced by the generalized permeability $P(|\varphi|)$, or by the function  $K(|\varphi|)$, naturally appear in the supersymmetric context. Vortex solutions satisfying the first order equations and saturating the inequality which relates the energy and the topological charge, would  then be candidates for the BPS configurations in supersymmetric extensions. The inclusion of fermions also contributes to the search of fermionic zero modes attached to the novel structures that we found in this work. This is also of current interest, and can be implemented under the lines of the recent work \cite{BM1}. Another issue concerns the extension of the $U(1)$ symmetry to the $U(1)\times U(1)$ case, to explore how the superconducting features that appeared before in Ref.~\cite{witten} could emerge in the context of the ringlike structures found in this work. The model with $U(1)\times U(1)$ symmetry can also be studied in the context of one visible and one hidden sector, under the lines of Ref.~\cite{hidden}.

The study of vortices in supersymmetry inspired models is a very active subject due to the so-called {\it web of dualities}, which generalizes the particle/vortex duality; see, e.g., Ref.~\cite{witten2}. It is expected that this duality has condensed matter applications, and an interesting possibility is to examine the particle/vortex duality in the case of vortices with internal structure. One can also examine the dynamics of vortices with internal structure in the presence of electric and magnetic impurities, to see how the internal structure modifies the scenario with standard vortices, investigated before in Refs.~\cite{tong,cuck}. Another issue of current interest concerns the presence of vortices in the gauged nonlinear Schr\"odinger equation and in nonlinear sigma models. These are interesting issues that can be implemented under the lines of Refs.~\cite{jackiw,nardelli}, to find magnetic vortices with local symmetry in order to advance towards condensed matter and make further contact with the recent work \cite{CM4} in which a phenomenological nonlinear sigma model is explored. This last issue is directly connected with the recent experimental observation \cite{expe} that suggests the presence of pair-density-wave excitations in the cuprate vortex halo, which is also examined in \cite{CM4,CM5} along distinct theoretical frameworks. 

\acknowledgements{The authors would like to thank the Brazilian agency CNPq and the Ministry of Education for partial support. DB thanks support from the grant CNPq:306614/2014-6, MAM thanks support from the grant CNPq:140735/2015-1 and DM thanks support from the grant No. 16-12-10344 of the Russian Science Foundation.}


\end{document}